\RequirePackage{fix-cm}
\documentclass[twocolumn]{svjour3}          
\smartqed  
\usepackage[utf8]{inputenc}
\usepackage{commath}
\usepackage{amsfonts}
\usepackage{amssymb}
\usepackage{color}
\usepackage{graphicx}
\usepackage[flushleft]{threeparttable}
\graphicspath{ {paper_folders/Images/} }
\usepackage{multirow}
\usepackage{listings}
\usepackage[linesnumbered,lined]{algorithm2e}

\newtheorem{defn}{Definition}[section]

\newcommand{\Sone}{{S_{r_{1}}}}
\newcommand{\Stwo}{{S_{r_{2}}}}
\newcommand{\Bone}{{B_{r_{1}}}}
\newcommand{\Btwo}{{B_{r_{2}}}}

\newcommand{\beq}{\begin{equation}}
\newcommand{\eeq}{\end{equation}}
\newcommand{\etal}{et al.}

\newcommand*{\affaddr}[1]{#1} 
\newcommand*{\affmark}[1][*]{\textsuperscript{#1}}

\begin{document}

\title{Robust Trajectory-based Density Estimation for Geometric Structure Recovery: Theory and Applications
}


\author{Turner Richmond\affmark[1]         \and
        Namita Lokare\affmark[1]           \and
        Qian Ge\affmark[1]                 \and
        Edgar Lobaton\affmark[1]
}


\institute{E. Lobaton \at
              Box 7911, NC State University, Raleigh, NC 27695, USA \\
              Tel.: +1-919-515-5151\\
              Fax: +1-919-515-5523\\
              \email{edgar.lobaton@ncsu.edu}    \\
            \affaddr{\affmark[1] Department of Electrical and Computer Engineering, North Carolina State University, Raleigh, NC 27695, USA}\\
}

\date{Received: date / Accepted: date}

\maketitle

\begin{abstract}
With the rise of the Internet of Things, strategies for effectively processing big data are essential for discovering meaningful insights. The time series datasets produced by groups of interconnected devices contain valuable underlying patterns. Recent works have extracted patterns from spatio-temporal datasets to aid in road network generation, activity recognition, and others. The speed and accuracy of the underlying geometry reconstruction are important in these applications. Existing methods such as kernel density estimation (KDE) have been used but are often computationally expensive. We propose modifying edge quadtrees to utilize their effective hierarchical structure. Our modification estimates density using a novel trajectory count function which provides mathematical guarantees on the stability of the count by enforcing an invariance to local perturbations. We evaluate our method’s effectiveness at extracting the underlying geometry and representative subsample points. For verification, we compare against a KDE variant at extracting the underlying shape of noisy synthetic trajectories travelling along the shape. We compare map extraction from GPS traces against current methods. Our method significantly improves runtime while extracting the geometry better or at least comparably. We also compare against maxmin subsampling on an activity recognition data set and find a significant runtime improvement with comparable performance.
\end{abstract}

\section{Introduction}\label{sec:introduction}


The increasing availability of large data sets containing spatio-temporal data is driving the need for data analysis methodologies. Spatio-temporal data has been widely used in various applications: traffic monitoring \cite{4059186,li2007traffic,li2006coarse}, trajectory similarity search used in route searching and semantic understanding, 
\cite{Chentrajectories,Ahmed:2014:LPH:2666310.2666390,Chen:2005:RFS:1066157.1066213,wang2006learning}, weather and pollution monitoring \cite{MAKRA20112630}, and anomaly detection \cite{CHAKER2017266,chandola2009anomaly,mahadevan2010anomaly}.
 
\begin{figure*}[!t]
	\centering
	\includegraphics[width=2\columnwidth]{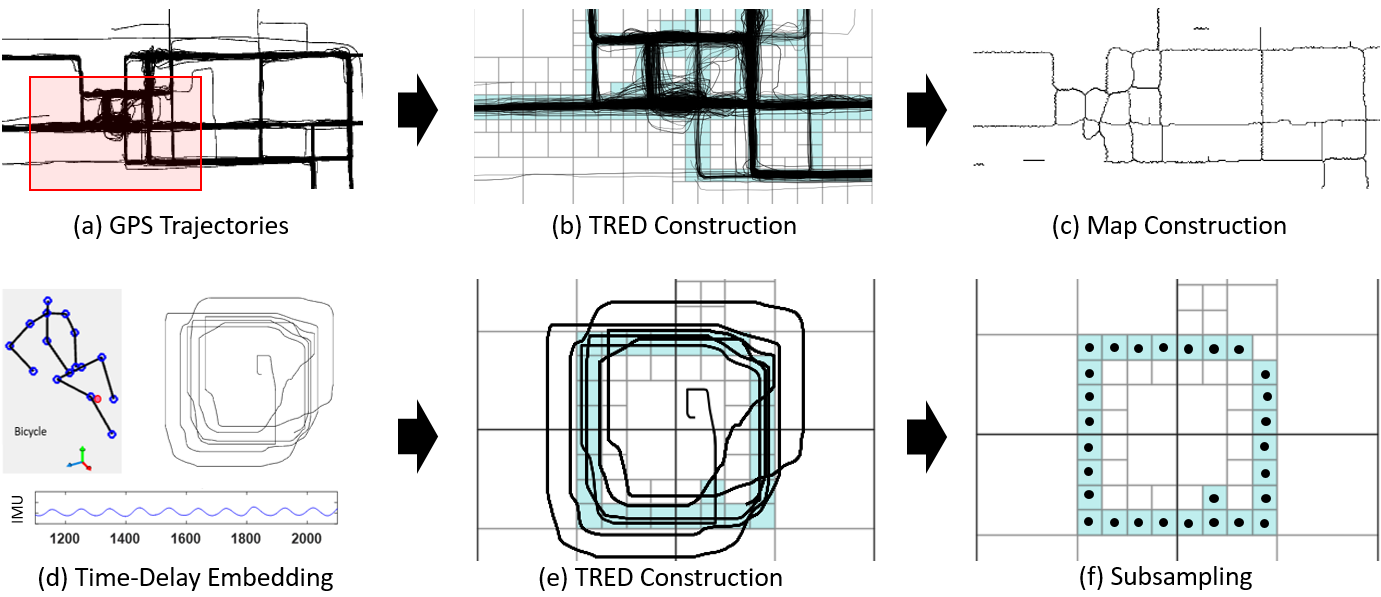}
    \caption{
    Pipelines for Map Reconstruction and Activity Recognition Applications. Along the top, we apply our method to raw GPS trajectories of Chicago (a) and show the hierarchical structure (b) of the red highlighted region. The skeleton of the reconstructed map from our method is shown in (c).  We show our method applied to an activity recognition pipeline which performs Time Delay Embedding (TDE) on inertial signals of a wearable devices (d). Our hierarchical structure (e) and subsampled points (f) are used to analyze the shape of the TDE for activity recognition purposes. 
    }
    \label{fig:Overview}
\end{figure*}

Trajectory clustering is an efficient method to analyze trajectory data. 
By counting the number of trajectories passing through regions of the data space, both the spread and density of the data can be evaluated.
Researchers have explored applying the principles of clustering trajectories to determine animal movement patterns \cite{wisdom2004spatial,BSNTC,Li2010}, mapping roads from vehicle GPS data \cite{Chen2010,Aanjaneya2011,Li2010}, and describing environmental characteristics such as fault lines \cite{Aanjaneya2011}, and hurricane forecasts \cite{powell2001accuracy}.

Trajectory clustering techniques are often used to supply a set of descriptive trajectories, however this information can also be obtained by learning the region's trajectory density. Density based methods have been used in the context of clustering as in DBSCAN \cite{DBSCAN}, where density of points has been taken into account for purpose of clustering. The density based method TRACLUS \cite{TRACLUS} is used to cluster segments of larger trajectories. Similarly, in \cite{Lobaton2016}, the authors use density of edges accumulated by aggregating segmentations by an ensemble scheme. 
Each of these practices has drawbacks associated with finding the density of trajectories in an efficient manner. In this paper, we explore the issues associated with using a point clustering algorithm, and aim to present an algorithm which is not highly sensitive to input parameters or trajectories, but is instead deterministic and robust.
In \cite{TRACLUS}, it is noted that the TRACLUS algorithm is sensitive to input trajectories where short input trajectories may produce undesirable clustering results.
Many of the density estimation approaches have large computational complexity which we aim to reduce.


Evaluating the shape of a density estimation can be computationally expensive.
Therefore in real-time application, evaluation of the geometry of a data set often requires subsampling the data.
Subsampling methods aim to downsample the data to keep the most representative data points while removing outliers \cite{Zimek:2013:SEE:2487575.2487676,de2004topological}.
A k-nearest neighbors density based subsampling together with the maxmin landmark selection algorithm (KNN-maxmin) is utilized in \cite{de2004topological}.
Many devices in real-time applications cannot guarantee a set sampling frequency and point-based methods fail in these cases because if we only consider densities of points we will end up with more samples in the segments where the sampling frequency was higher. 
The need to minimize the affect of sampling rate on the density of paths is another motivation for the work present in this paper.
We compare our proposed method to the maxmin subsampling algorithm \cite{de2004topological} to highlight these issues and also to validate our method.

The strategy presented in this paper aims to generate a level set density function of time series data sets.
We can utilize the density function to subsample points in efficient manner which captures the structure of the data. 
By utilizing the density function, we are able to uniformly subsample the entire structure of the data without costly pointwise distance comparisons.  Furthermore, the sampling frequency does not affect the shape of the density function and thus our process is robust to changes in sampling frequency along a trajectory while point based subsampling may be affected by uneven sampling.

As an example application, activity recognition is a common problem in wearable sensing applications. The authors of \cite{alireza_TDE} show that the structure of the person's joint data can be recovered by topological feature extraction. This method allows for real-time activity recognition however, suffers from the issues of point-based subsampling method. We improve on this method by employing our trajectory based subsampling strategy and speed the process by $75\%$ while achieving similar performance.

We provide an offline as well as an update implementation \footnote{Code available at: https://zenodo.org/record/2587033} of our existing geometric structure.
Offline processing can be utilized in the case of large data sets or for performing exploratory data structure analysis as in the case of generating an initial road map from GPS trajectories, see (a-c) in Figure \ref{fig:Overview}, and it can be updated once new trajectory becomes available.
Real time data applications can benefit from updating an existing or expected hierarchical data structure representing the underlying geometry, see (d-f) in Figure \ref{fig:Overview}.


The rest of this article is organized as follows; we present the problem formulation as well as our density function in Section \ref{sec:formulation}, followed by the implementation details in Section \ref{sec:implementation}. Section \ref{sec:synthetic_data} describes the synthetic data generation and evaluation of recovering the underlying shape from data with harmonic and impulse noise. Section \ref{sec:activity_recognition} presents the activity recognition data set, pre-processing steps, and the evaluation of our method on the data set. Section \ref{sec:street_map} presents the application of our method on Street Map generation. We conclude in Section \ref{sec:conclusion} to highlight the benefits and limitations presented in the paper.
\section{Mathematical Formulation}
\label{sec:formulation}

\begin{figure}[!b]
	\centering
	\includegraphics[width=\columnwidth]{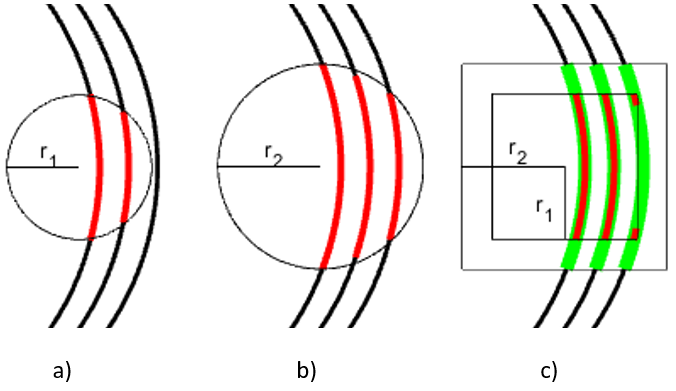}
    \caption{Illustration of count function. The red portions of the trajectories are the elements of sets $A_{B_{r_1}}$, $A_{B_{r_2}}$, and $A_{S_{r_1}}$. The green portions of the trajectories are the elements of the set $A_{S_{r_2}}$. As shown $C_{B_{r_1}}=2$, $C_{B_{r_2}}=3$, $C_{S_{r_1}}=4$ and $C_{S_{r_1},S_{r_2}}=3$. As specified in Theorem \ref{thm:boundedCt} $C_{B_{r_1}}\leq C_{S_{r_1},S_{r_2}} \leq C_{B_{r_2}}$.}
    \label{fig:Counts}
\end{figure}

Given a set of trajectories in $\mathbb{R}^d$, our objective is to characterize how their density varies across the space. We will characterize the density by counting the number of trajectory segments in a neighborhood of radius $r \in \mathbb{R}^+$ at a location $x \in \mathbb{R}^d$ in the space with coordinates ($x_1, \ldots ,x_d$). This defines a local count function over the space. For our analysis, we aim to (1) develop a computationally efficient way to estimate these counts over the entire space, (2) show that these functions are stable, and (3) define a sampling strategy that produces points that have a similar structure to that of a specific level set of the count function. As we will see later, we will analyze and compare the structure of a level set by characterizing the local topological structure.

In order to formalize the problem, we introduce some notation. In particular, for a set ${\{\gamma_{k}\}^{N}_{k=1}}$ of continuous trajectories, we define $\gamma_{k} : [0,T_{k}] \rightarrow \mathbb{R}^{d}$ and their trace over an interval $I$ as  $\gamma_{k}(I) = \{ \gamma_{k}(t) \; | \; t \in I \} \subset \mathbb{R}^d$. For simplicity, the analysis will focus on a single trajectory (i.e., we drop the index $k$ in $\gamma_k$) in $\mathbb{R}^2$ that has continuous second derivatives, but the definition and results can be directly generalized to multiple trajectories and higher dimensions.

In order to define our local count function, we will make use of the closed disk $B_r(x) = \{ y \in \mathbb{R}^2 \; | \; (y_1-x_1)^2 + (y_2-x_2)^2 \leq r^2 \}$. We consider the pair of entry / exit points 
\beq 
A_r(x) = \left\{ [a_i,b_i] \; | \; \gamma([a_i,b_i])\subset B_r(x) \textrm{ is maximal}\right\},
\eeq
where by maximal we mean that there is no larger interval $[a,b] \subset [0,T]$ which contains $[a_i,b_i]$ with trace fully contained in $B_r(x)$. Formally, we define the \emph{local trajectory count function} $C_{B_{r}} : \mathbb{R}^{2} \rightarrow \mathbb{N}$ as:
\begin{equation} \label{eqn:DiskCount}
    C_{B_{r}}(x) = \abs{A_r(x)},
\end{equation}
where $\abs{A}$ is the cardinality of set $A$. Figure \ref{fig:Counts} (a) shows an example of the segments obtained as part of the set $A_r(x)$.

\subsection{Approximating the Local Count Function}
\label{sec:approximation}

Directly computing $C_{B_{r}}(x)$ over multiple scale parameters $r$ would require determining the segments of the trajectories within a given region ($B_{r}$) for every location ($x$) in the space. This can be computationally intensive. Hence, it would be beneficial to make use of hierarchical data structures such as quadtrees, which can partition trajectories into segments at a coarse scale and then subpartition these segments when transitioning to a finer scale. However, quadtrees partition the space into square regions, so we require a count based on square neighborhoods (instead of disks) in order to exploit the computational efficiency of the representation. Figure \ref{fig:Overview} (top) illustrates how to use a hierarchical structure that allows the efficient computations of these counts.

Let us define the closed square $S_r(x) = \{ y \in \mathbb{R}^2 \; | \; \max(\abs{y_1-x_1},\abs{y_2-x_2)}) \leq r \}$. We can define a count $C_{S_r}(x)$ in a similar way as we did in Equation \ref{eqn:DiskCount}. However, as we will see in section \ref{sec:stability}, this function is sensitive to small perturbations of the trajectories, which can make the counts arbitrarily large. Hence, we define a \emph{robust local square count function}
\begin{equation}
    C_{{S_{r_{1}}},{S_{r_{2}}}}(x) = \abs{ A_{S_{r_{1}}}(x) / \sim}
\end{equation}
where $r_{1}<r_{2}$ and $A_{S_{r_{1}}} / \sim$ is the equivalence class corresponding to the set of intervals in $A_{S_{r_1}}$ identified under the equivalence relation $\sim$. A pair of intervals $[a_1,b_1]$ and $[a_2,b_2]$ in $A_{S_{r_1}}$ are equivalent (i.e., $[a_1,b_1] \sim [a_2,b_2]$) if there exists an interval $[a,b] \in A_{S_{r_2}}$ such that $[a_1,b_1] \cup [a_2,b_2] \subset [a,b]$. Figure \ref{fig:Counts} illustrates this square count and its relationship to the disk counts.

We can guarantee, given a bound on curvature, that the robust local square count function approximates the disk count as it is shown in the theorem below. The proof of this result is outlined in the Appendix \ref{sec:curv_proof}.

\begin{theorem} \label{thm:boundedCt}
Given that $\sqrt{2} \cdot r_{1} \leq r_{2} < \frac{1}{\kappa_{max}}$, where $\kappa_{max}$ is a bound on the maximum curvature of a $\gamma$, then for all $x \in \mathbb{R}^2$ we have that
\begin{equation} \label{eq:thm1}
    C_{\Bone}(x) \leq C_{\Sone,\Stwo}(x) \leq C_{\Btwo}(x).
\end{equation}
\end{theorem}

\subsection{Stability of Local Count Functions}
\label{sec:stability}

When trajectory densities are counted, it is important to define a notion of stability of the count in a region such that the change in a region's count is bounded when a small $\epsilon$-perturbation is applied. We begin by formalizing our notion of stability.

\begin{defn}
An {\bf $\epsilon$-perturbation} to a trajectory $\gamma$ with maximum curvature bounded by $\kappa_{max}$ produces a new trajectory $\hat{\gamma}$ with the same maximum curvature bound such that $||\hat{\gamma}(t)-\gamma(t)||_2 \leq \epsilon$. We say that the $\epsilon$-perturbation is {\bf small} if $\epsilon < \frac{1}{\kappa_{max}}$.
\end{defn}

\begin{defn}
A count function is {\bf stable} if for any trajectory $\gamma$ with maximum curvature bounded by $\kappa_{max}$ the change in its value due to a small $\epsilon$-perturbation is bounded everywhere. The bound may depend on the actual trajectory. Otherwise, the count is said to be {\bf unstable}.
\end{defn}

It may seem that a bound depending on $\gamma$ may seem too relaxed of a condition. However as can be seen from the proof of the theorem below, a bound on the change of the count $C_{B_r}$ that is independent of $\gamma$ can be found if we constrain our definition to curves of fixed length.


\begin{figure}
	\centering
	\includegraphics[width=\columnwidth]{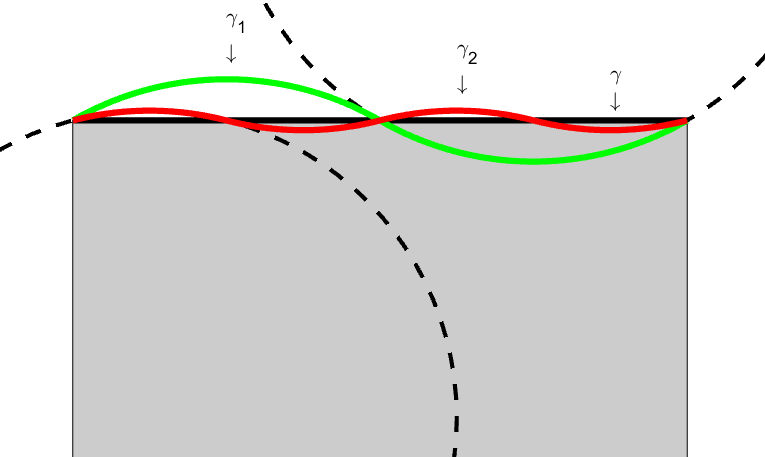}
    \caption{Illustration of unstability of square count. Let $\gamma$ be a trajectory (shown in black) that moves along the boundary of the square $S_r$ (shaded in gray). The count for this trajectory is 1. We can create a sequence of $\epsilon$-perturbations such that the count function is not bounded. An initial trajectory can be constructed by perturbing the trajectory to match a number of arcs of circles with radius $r>\frac{1}{\kappa_{max}}$ (as shown in green). The count of $\gamma_1$ is 2 (counting the left end point). By construction, the perturbation will be small. We can create a new trajectory (shown in red) by increasing the number of arcs to be twice as many. The count of $\gamma_2$ is 3. This process can be repeated indefinitely.}
    \label{fig:UnstableSquareCount}
\end{figure}

The following theorem shows that indeed the count over a disk neighborhood is stable, and the count over a square unstable while the robust square count is stable. Figure \ref{fig:UnstableSquareCount} provides an example in which an $\epsilon$-perturbation can be constructed to cause an unbounded increase on the square count. This may seem like an unlikely scenario. However, we observed situations in our experimental validation in which the square counts were much larger than the disk counts, which corresponded to scenarios similar to the one in the figure. Proof of this theorem is outlined in the Appendix \ref{sec:stab_proof}.

\begin{theorem} \label{thm:stability}
$C_{S_{r}}$ is an unstable count. $C_{B_{r}}$ is stable given that $r < \frac{1}{\kappa_{max}}$. $C_{\Sone , \Stwo}$ is stable to perturbations of a trajectory given that $r_2 < \frac{1}{\kappa_{max}}$.
\end{theorem}

\subsection{Density-based Sampling}
 
Our method also provides a simple choice for subsample points when a small set of geometrically representative points are required for shape analysis.
As it will be described in the next section, a hierarchical representation based on a quad-tree structure is constructed as Figure \ref{fig:Overview} illustrates. At the coarser scale, segments of a trajectory within a given square region will be extracted. Then, the region will be further split into smaller squares, and the process will be repeated. The number of segments at each scale will be counted using our robust local square counting scheme. The centers of squares at the finest scale with a count higher than specified threshold (i.e., regions with high enough density) will be taken as samples from the targeted superlevel set of the trajectory density function. We refer to this scheme as the \emph{Trajectory-based Representation for Estimation of Density} (TRED) representation.

\section{Implementation Details}
\label{sec:implementation}



\begin{algorithm}[!b]
\SetAlgoLined
\KwIn{$\gamma$, $\delta_r$, $\tau$, $M$, $R$}
\KwOut{$\mathcal{C}$,$\mathcal{S}$,$\Lambda$}
 \tcp{Initializing segments and counts}
 $\mathcal{S}_{0,1} = \gamma$\;
 $\mathcal{C}_{0,1} = 1$\;
 \tcp{Initializing list of active bins}
 $\Lambda_0 = \{1\}$\;
 \tcp{Iterating over refinement levels}
 \For{$m \in \{1, \cdots, M\}$}{
  $\Lambda_m = \{\}$\;
  \tcp{Defining radii for robust count}
  $r_{1m} = R\cdot 2^{-m}$\;
  $r_{2m} = r_{1m} + \delta_r$\;
  \tcp{Iterating over active bins}
  \For{$k \in \Lambda_{m-1}$ } {
   \tcp{Iterating over children of bin}
   \For{$p \in Children_{m-1,k}$} {
    \tcp{Computing counts and segments}
    $[\mathcal{S}_{m,p},\mathcal{C}_{m,p}] = F(S_{m-1,k},x_{m,p},r_{1m},r_{2m})$\;
    \tcp{Updating active bin list}
    \If{$C_{m,p} > \tau$} {
     $\Lambda_m.append(p)$
    }
   }
  }
 }
\caption{TRED Offline Pseudocode}
\label{alg:TRED}
\end{algorithm}

For our implementation of TRED, we use the leaf-unbalanced quadtree data structure introduced by Finkel and Bentley \cite{Finkel} to refine a search space in order to extract the underlying geometric structure of the data set.
The generalizable nature of the quadtree allows for it to be run in higher dimensions with the need to define only a few parameters.
We refer to the nodes of the quadtree as bins which are always square regions. Algorithms \ref{alg:TRED} and \ref{alg:TRED_ins} 
provide pseudocode. The parameters required are:
\begin{enumerate}
  \item The {\bf threshold $\tau$} used for specifying a stop criterion for splitting the bins. If we want to compute the counts for all scales then we can select $\tau=0$. 
  \item A {\bf length for the base square $R$} that encloses the entire trajectory. It specifies the initial region to be split.
  \item A {\bf maximum depth $M$} of the quadtree. We can select this depth such that at the finest scale the square size $r_1 = R\cdot 2^{-M}$ satisfies any assumptions about our curvature constraints.
  \item A {\bf radius offset $\delta_r(\cdot)$} used to specify the difference between $r_1$ and $r_2$ at every scale. If we assume that the finest scale satisfies the curvature constraints then a good choice for this offset is a constant  $\delta_r = (1-\sqrt{2})\cdot R \cdot 2^{-M}$.
\end{enumerate}

Let us review the parameters required for a successful application of TRED.
Trajectory threshold defines how many times a set of trajectories needs to pass through a region before it should be split into child bins.
Thus threshold indicates the density required to be included in a resulting superlevel set, where higher thresholds produce structures with more restricted superlevel sets.
It follows that the density of the trajectories is the most influential factor in choosing an effective threshold.
The choice of bin depth defines at what point a bin no longer splits into child bins after it exceeds the trajectory threshold.
Bin depth is closely coupled to maximum curvature and the size of the region as the finest bin side is a function of the region size and the maximum bin depth.
Thus treating the depth accordingly, the region occupied by the trajectories along with the maximum curvature and desired resolution of the space can be used to estimate an effective bin depth.
We expect that larger bins will include higher counts of trajectories as compared to the children of the bin. As a direct result, modifying the bin depth or region size may affect the optimal choice of trajectory threshold.
Once the first three parameters are known, the value for radius offset must be decided.
We have shown the importance of defining maximum curvature in Section \ref{sec:approximation}, and as such it is vital to choose an effective offset.
Estimated maximum curvature should be used to determine the radius offset once the region size and maximum depth are known utilizing the aforementioned formula for $\delta_r$. Of course, from a data-driven approach, we can consider all these variables as hyper-parameters to tune.

\noindent {\bf Offline Version.} Algorithm \ref{alg:TRED} stores all segments in the list $\mathcal{S}$. These segments correspond to the squares $S_{r_{2m}}$ used in the robust count computation. The list $\Lambda$ is used to keep track of the active bins to be refined at each scale based on the count criteria used in line 11. For each scale and active bin, counts and corresponding segments for the children are computed. Line 10 calls a function that returns the segments associated with $S_{r_{2m}}(x_{m,p})$ and the counts associated with $C_{S_{r_{1m}},S_{r_{2m}}}(x_{m,p})$, where $x_{m,p}$ is the center of the bin at scale $m$ and index $p$. The inputs for this function are the segments in the parent bin, center for the child bin, and the corresponding radii. Finally, line 12 updates the list of active bins at the corresponding scale given that the count criterion is satisfied.
The algorithm returns the lists containing counts $\mathcal{C}$, segments $\mathcal{S}$, and active bins $\Lambda$.

\begin{algorithm}[!t]
\SetAlgoLined
\KwIn{$\gamma'$,$\mathcal{C}$,$\mathcal{S}$,$\Lambda$,$\delta_r$,$\tau$,$M$,$R$}
\KwOut{$\mathcal{C}$,$\mathcal{S}$,$\Lambda$}
 $\hat{\mathcal{S}}_{0,1} = \gamma'$\;
 $\mathcal{S}_{0,1}.append(\gamma')$\;
 $\mathcal{C}_{0,1} = {C}_{0,1} + 1$\;
 \For{$m \in \{1, \cdots, M\}$}{
 $r_1 = R\cdot 2^{-m}$\;
 $r_2 = r_1 + \delta_r$\;
  \For{$k \in \Lambda_{m-1}$ } {
   \For{$p \in Children_{m-1,k}$} {
    $[\hat{\mathcal{S}}_{m,p},\hat{\mathcal{C}}_{m,p}] = F(\hat{S}_{m-1,k},x_{m,p},r_{1m},r_{2m})$\;
    $\mathcal{S}_{m,p}.append( \hat{\mathcal{S}}_{m,p})$\;
    $\mathcal{C}_{m,p} = \mathcal{C}_{m,p} + \hat{\mathcal{C}}_{m,p}$\;
    \If{$p \not\in \Lambda_m$ and $\mathcal{C}_{m,p} \geq \tau$} {
     $\Lambda_m.append(p)$\;
     $\hat{\mathcal{S}}_{m,p} = \mathcal{S}_{m,p}$\;
    }
   }
  }
 }
 \caption{TRED Update Pseudocode}
\label{alg:TRED_ins}
\end{algorithm}


\vspace{0.05in}
\noindent {\bf Update Version.} Since we are also considering applications in which trajectory data will be incrementally provided to us (e.g., one trajectory at the time after a vehicle completes a trip for road reconstruction), we also provide an update version of the algorithm that can update the counts by iterating over consecutive samples over time (see Algorithm \ref{alg:TRED_ins}). Additionally, we ambition streaming versions of the algorithm for applications such as activity recognition in which data points could be added and removed over time.


\vspace{0.05in}
In the following sections, we run the algorithms presented using a computer with 16 GB of RAM and an Intel i7-3770 CPU at 3.4 GHz. TRED was implemented using python 3.6 and compared against existing implementations of algorithms. The algorithms used for comparison in Sections \ref{sec:synthetic_data} and \ref{sec:activity_recognition} are implemented in Matlab while the algorithms used for comparison in Section \ref{sec:street_map} are implemented in a variety of languages including python, Matlab, and Java.


        
\section{Synthetic Data Evaluation}
\label{sec:synthetic_data}

We utilize synthetic data to evaluate the ability of TRED to recover an underlying geometric shape given noisy data.
TRED is evaluated against a brute force trajectory local density estimation.
The brute force method calculates $C_{B_{r}}(x)$ for points in a dense regular grid.
We will refer to our brute force method as Trajectory-based Local Density Estimation (TLDE) for ease of reference.
We note that TLDE may be compared to a kernel density estimation (KDE) with a uniform kernel followed by a threshold applied to generate a binary grid.
The comparison aims to identify local topological differences between the ground truth and the two geometric recovery techniques (TRED and TLDE).
Below, we provide details on the data generation followed by the evaluation. 

\subsection{Data Generation}

\begin{figure}[!b]
	\centering
	\includegraphics[width=\columnwidth]{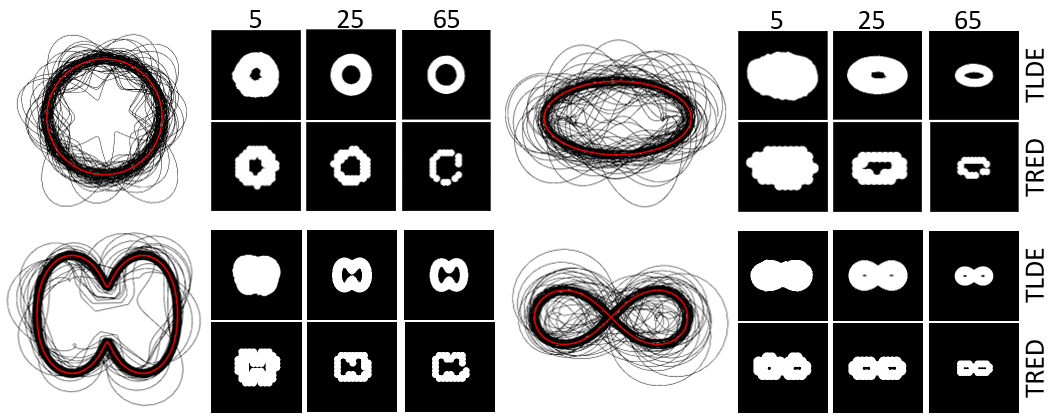}
    \caption{
    Four synthetic trajectories with perturbations (black) with the reference trajectory (red) for each shape. Next to the shapes, we show TRED and TLDE recoveries using the thresholds 5, 25, and 65. Note that the ground truth is often very similar to TLDE at a threshold of 25. 
    }
    \label{fig:Synthetic_raw}
\end{figure}
We evaluate the performance using four synthetic shapes shown in Figure \ref{fig:Synthetic_raw}.
The circle was generated with a radius of $1$. The ellipse was generated with a semi-major axis of $1$ and a semi-minor axis of $0.5$. The ``Eight" is a ``Lemniscate of Bernoulli" with width $1$. The ``Peanut" is a ``Cassini Oval" with distances between the centers set to $0.92$, product of distances from centers set to $1$, and then the x-axis is scaled such that its range is $[-1,1]$.
A reference trajectory is generated by creating the shape without any perturbations. 

For each shape, 200 samples were generated where each sample contains 100 cycles of the shape.
Each sample contains both harmonic and random noise added perpendicularly to the direction of travel.
The magnitude of the harmonic noise for each sample is selected using a uniform random sampling from $0.05$ to $0.1$ for the peanut and $0.02$ to $0.07$ for the circle, eight, and ellipse.
The random noise is additionally added to the harmonic noise. The random noise for each sample has between $10$ and $80$ pulses with magnitude between $0.1$ and $0.5$ where the quantity and magnitude of pulses is selected using a uniform random sampling. 


\subsection{Evaluation}


For TRED, we set $R=6$ for all shapes and use a maximum depth $M=5$. For a radius of $r_1 = R\cdot 2^{-M}$, if the count function $C_{S_{r_1,r_2}}(x)$ is greater than the given threshold then all locations within $B_{r_2}(x)$ are set to 1.
Similarly for TLDE with $r=0.3$, all locations within $B_{r}(x)$ are set to 1 if $C_{B_{r}}(x)$ is greater than the given threshold.
Given these mappings over the dense regular grid containing the trajectories, we can compute the local homology differences between TRED and the ground truth as well as between TLDE and the ground truth.

The level sets produced for both TRED and TLDE as described above are compared to the ground truth level set for each shape.
The ground truth level sets are produced using TLDE with a threshold of 1 run on the reference trajectory for the shape. Note the reference trajectory has no noise, so the count for TLDE is $C_{B_{r}}(x) \in \{0,\; 100\}$.
When comparing the level set of TRED or TLDE against the ground truth, we use a topologically aware metric presented by Ge et a. \cite{Ge2020} with a Lipschitz deformation bound 
$K_d = 0.096$
and the radius of the region $\rho = 0.48$. The parameters for the metric are selected to quantify what a meaningful local homology difference is, and for this reason different parameters could be selected to relax the metric.



\begin{figure}[ht]
	\centering
	\includegraphics[width=\columnwidth]{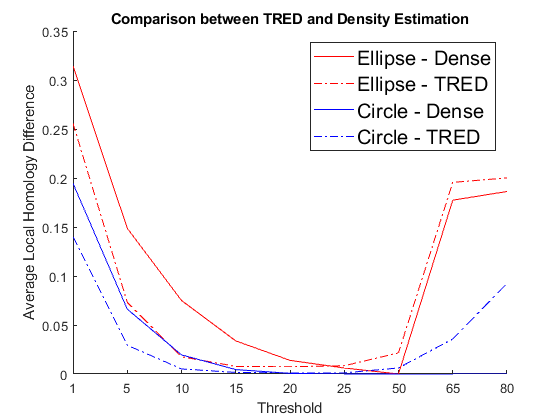}
    \caption{Comparison of local homology difference between TRED and the local density estimation of trajectories at different count thresholds. 
    }
    \label{fig:Synthetic_Results}
\end{figure}

\begin{figure}[ht]
	\centering
	\includegraphics[width=\columnwidth]{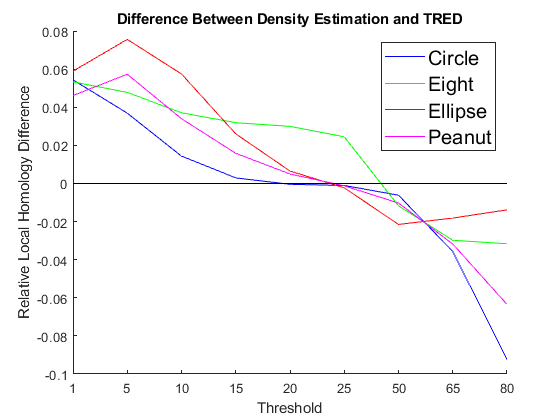}
    \caption{TRED average local homology difference subtracted from TLDE. As shown, TRED performs better at lower thresholds, and TLDE performs better at higher thresholds.
    }
    \label{fig:Synthetic_Results_Relative}
\end{figure}

The results of both TRED and TLDE compared against the ground truth are shown in Figure \ref{fig:Synthetic_Results}. As expected, the threshold needs to be increased in order to reduce the effect of noise on the resulting level set. 
We use Figure \ref{fig:Synthetic_Results_Relative} to show how effective TRED and TLDE are at extracting the underlying geometry across all shapes. The values plotted indicate the difference between the values shown in Figure \ref{fig:Synthetic_Results} at each threshold for each shape.

We expect the local density estimation to improve rapidly, and produce level sets with the same local topology structure as the reference signal as the perturbations are filtered out due to the higher threshold.
Due to the small number of pulse perturbations with a large magnitude, the metric improves more rapidly at smaller thresholds.
Yet as the threshold increases above 50, we see a large local homology difference due to the decreasing width of the level set. In Figure \ref{fig:Synthetic_raw}, the level sets with a threshold of 25 have low local homology difference while level sets with a threshold of 65 have a large local homology difference. Despite the same global topology, these level sets have too great a difference in their widths to have a similar local topology.
Note the gaps in TRED mask with a threshold of 65. Due to the high threshold and the granularity of the data structure, some gaps appear which are not present in the higher resolution TLDE. For this reason, it is important to choose an effective threshold for a particular depth.


As shown in Figure \ref{fig:Synthetic_Results}, we see TRED and TLDE have similar trends where the local homology difference improves until the threshold imposes too strict a constraint on level set.
Yet, overall we see that TRED is able to outperform TLDE when extracting the geometric structure of a shape.
As TLDE increases the threshold, it approaches the ground truth until the threshold surpasses 50 at which point the level set is too thin to be considered locally similar.
Yet even at the optimal TLDE thresholds, TRED is able to perform comparably.
The two strategies do not have the same optimal thresholds, and as such it is important to find the best parameters for TRED using experimental validation and application knowledge. 

To conclude, we consider the run time complexity to compare these two approaches. TRED has an $O(n)$ runtime on the synthetic dataset while TLDE has a runtime complexity of $O(nm)$ where $m$ is the number of locations at which $C_{r_1}(x)$ is calculated. For this experiment $m=500^2$ which effectively squares the runtime. At lower values of $m$ we would expect to see an improved runtime, but the resulting level set would have lower resolution. When the resolutions between TRED and TLDE are comparable, TRED still maintains a superior run time complexity due to the hierarchical nature of the data structure.

\section{Activity Recognition Evaluation}
\label{sec:activity_recognition}

In this section, we show how TRED can be utilized for activity recognition. Our goal is to show that TRED performs comparably or even better than the existing subsampling strategy (maxmin) while achieving a significant gain in computational speed.

\subsection{Activity Recognition Data Set}
\label{sec:ActivityDataset}

\begin{figure*}[!t]
	\centering
	\includegraphics[width=2\columnwidth]{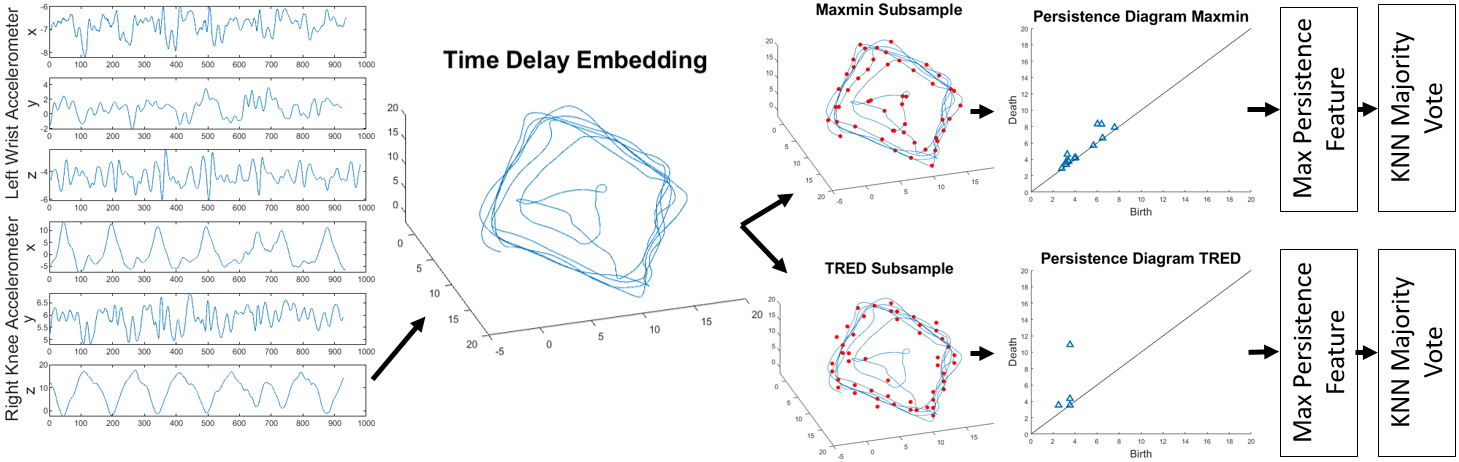}
    \caption{
    Pipeline for activity recognition where Time Delay Embeddings (TDE) (one for each acceleration channel) are computed. We show the pipeline for the two different subsampling techniques. The KNN majority vote is then performed using the 6-Dimensional topological features extracted from each TDE.
    }
    \label{fig:ActivityOverview}
\end{figure*}

The NCSU-ADL data set presented by Lokare \etal \ \cite{namitaInventions} is used for real-time activity recognition.
The data set contains physiological and motion data for daily living activities from $11$ healthy individuals. 
From the data set, the 3-axis accelerometer measurements of the devices on an individual's left wrist and right leg are utilized to classify activity.
Single windows of $1000$ and $2000$ samples are of interest which amount to $5$ and $10$ seconds respectively.
Table \ref{tab:ADLDesc} describes the activities of interest in the NCSU-ADL data set categorized by activity type, activity frequency, and sampling frequency. Activity frequency is extracted using the first zero crossing of the autocorrelation function. This is averaged over all the subjects.The activity frequency gives us the time period, on average, needed for a subject to complete a full motion for a particular activity.

\begin{table}[!b]
\centering
\caption{NCSU-ADL Data Description
}
\label{tab:ADLDesc}
{\renewcommand{\arraystretch}{1.3}
\begin{tabular}{|l|c|c|c|}
\hline
Activity                 & Activity Type                     & \begin{tabular}[c]{@{}l@{}}Activity \\ Frequency \end{tabular}            & \begin{tabular}[c]{@{}l@{}}Sampling \\ Frequency\end{tabular}                     
\\ \hline \hline
Bicycling              & Periodic     & 1.16 Hz    & 202 Hz      
\\ \hline
Rowing                 & Periodic     & 0.40 Hz    & 202 Hz      
\\ \hline
Walking                & Periodic     & 1.42 Hz    & 202 Hz      
\\ \hline
Carrying A Box         & Semi-Periodic      & -          & 202 Hz      
\\ \hline
\end{tabular}} \quad
\end{table}

The NCSU-ADL data set is processed using the method developed by Dirafzoon \etal \ \cite{alireza_TDE}.
The accelerometer data is windowed over time into the 1000 and 2000 sample windows previously mentioned. Transitions of activity (windows containing multiple activities) are discarded from the study since they are very noisy even after filtering. We perform a 4-fold cross validation scheme over the entire data set where each subject is equally represented in each partition of data.

The activity classification algorithm used in \cite{alireza_TDE,EUSIPCO} 
is shown in Figure \ref{fig:ActivityOverview} and is summarized as follows:
\begin{itemize}
\item Thread the x, y, and z accelerometer from the right knee and left wrist as features. For each one of the six features, perform time delay embedding (to a three dimensional space) using the specified windows (1000 or 2000 samples).
\item Subsample the embedding utilizing either maxmin subsampling or TRED.
\item Extract a feature based on the topologically persistent hole for each feature using Persistent Topology.
\item Classify the window using a majority vote over the k-nearest neighbors.
\end{itemize}
In this article we expand on the analysis presented in \cite{EUSIPCO} for the above pipeline.

\subsection{Activity Recognition Evaluation}

The activity prediction method used in \cite{alireza_TDE} is applied on the NCSU-ADL data set as described in Section \ref{sec:ActivityDataset}. We compare our method TRED against the maxmin subsampling method \cite{de2004topological}. Two window sizes of 1000 and 2000 samples which correspond to 5 and 10 seconds, respectively, are chosen to compare the two methods.

\begin{figure}[h]
\centering
  \includegraphics[width=\columnwidth,keepaspectratio]{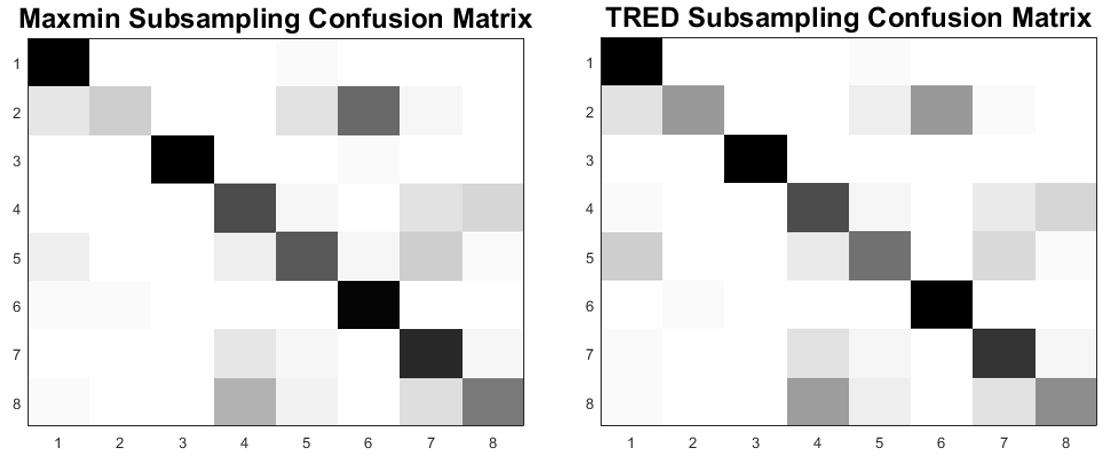}
\caption{Confusion Matrices for 5 Sec Windows. Activities are labeled as follows: (1) Row, (2) Carry Box, (3) Bicycle, (4) Rest, (5) Set Dinner, (6) Walk, (7) Type, (8) Lying.
}
\label{fig:confusion}
\end{figure}

Tables \ref{tab:f1subj} and \ref{tab:f1act} show the comparison of TRED against the maxmin subsampling strategy categorized by subject and activity, respectively. Periodic activities are of interest to us when using time delay embedding classification. We show the F1 score performance of both methods for non-periodic as well as complex activities for completion. Figure \ref{fig:confusion} shows the confusion matrix for both methods.

Overall we see that TRED outperforms maxmin subsampling in classification performance for periodic activities. Among the periodic activities, we see performance of TRED is lower for ``Rowing'' activity as compared to maxmin subsampling in case of the 5 sec window. Table \ref{tab:ADLDesc} shows that rowing has a long period (2.5 seconds) which results in generation of 2 cycles when time delay embedding is performed. As described above, it is important for trajectories to be clustered in high numbers to be successfully subsampled in TRED. Due to the long period of rowing combined with the inherent noise of the sensors, subsampling with TRED utilizing 5 seconds is prone to misclassification of windows where the waveform is perturbed either due to noise or inconsistency of a periodic motion.

\begin{table}[!b]
\centering
\caption{F1 Scores Categorized by Subject}
\label{tab:f1subj}
{\renewcommand{\arraystretch}{1.3}
\begin{tabular}{|c|c|c|c|c|}
\hline
                       & \multicolumn{2}{c|}{5 Sec Window}                 & \multicolumn{2}{l|}{10 Sec Window} \\ \hline
\multicolumn{1}{|l|}{Subject}  & \multicolumn{1}{l|}{TRED} & \multicolumn{1}{l|}{Maxmin} & TRED    & \multicolumn{1}{l|}{Maxmin}   \\ \hline \hline
1                      & {\textbf{0.861}}            & 0.844                       & 0.842   & 0.756                         \\ \hline
2                      & {\textbf{0.835}}            & 0.834                       & 0.789   & 0.754                         \\ \hline
3                      & 0.691                     & 0.640                       & {\textbf{0.695}}   & 0.642                         \\ \hline
4                      & 0.787                     & {\textbf{0.839}}                       & 0.739   & 0.760                         \\ \hline
5                      & 0.786                     & {\textbf{0.800}}                       & 0.736   & 0.689                         \\ \hline
6                      & 0.787                     & {\textbf{0.827}}                       & 0.720   & 0.747                         \\ \hline
7                      & 0.771                     & 0.760                       & {\textbf{0.784}}   & 0.658                         \\ \hline
8                      & {\textbf{0.786}}                     & 0.780                       & 0.752   & 0.763                         \\ \hline
9                      & {\textbf{0.767}}                     & 0.764                       & 0.730   & 0.743                         \\ \hline
10                     & 0.761                     & {\textbf{0.803}}                       & 0.775   & 0.786                         \\ \hline
11                     & 0.723                     & {\textbf{0.741}}                       & 0.709   & 0.678                         \\ \hline \hline
\textbf{Average}                & 0.777                     & {\textbf{0.785}}                       & 0.751   & 0.725  
 \\ \hline
\end{tabular}}\quad
\end{table}

\begin{table}[h]
\centering
\caption{F1 Scores Categorized by Activity Type 
}
\label{tab:f1act}
{\renewcommand{\arraystretch}{1.3}
\begin{tabular}{|l|c|c|c|c|}
\hline
                       & \multicolumn{2}{c|}{5 Sec Window}                 & \multicolumn{2}{c|}{10 Sec Window} \\ \hline
\multicolumn{1}{|l|}{Activity}  & \multicolumn{1}{c|}{TRED} & \multicolumn{1}{c|}{Maxmin} & TRED    & \multicolumn{1}{c|}{Maxmin}   \\ \hline \hline
Bicycling                & {\textbf{0.981}}            & 0.970                       & 0.930   & 0.931                         \\ \hline
Rowing                 & 0.891                     & {\textbf{0.926}}              & 0.887   & 0.836                         \\ \hline
Walking                   & {\textbf{0.923}}            & 0.880                       & 0.848   & 0.778                         \\ \hline
Carrying A Box            & {\textbf{0.480}}            & 0.280                       & 0.354   & 0.115                         \\ \hline
Seting Dinner             & 0.588                     & {\textbf{0.652}}              & 0.595   & 0.613                         \\ \hline
Typing                   & {\textbf{0.747}}            & 0.741                       & 0.728   & 0.735                         \\ \hline
Resting                   & 0.653                     & {\textbf{0.670}}              & 0.628   & 0.610                         \\ \hline
Lying                  & 0.492                     & {\textbf{0.555}}              & 0.495   & 0.469                         \\ \hline \hline
Periodic Activities                  & {\textbf{0.908}}                     & 0.897              & 0.861   & 0.818                         \\ \hline
\end{tabular}}\quad
\end{table}


We see significant improvement in performance of TRED over maxmin subsampling in the ``Carrying a Box'' and ``Walking'' activities. Differentiating ''Walking'' from the complex activity of ``Carrying a Box'' which contains the same movement in the lower body is a notable advantage to utilizing the TRED subsampling method. We see the improvement due to TRED's noise filtering property where single large variations in the time delay embedding are no longer considered representative sections of the embedding. The filtering of noise is most notably seen in the wrist. Walking has a consistent movement with a clear structure in the time delay embedding, yet the wrist movement in carrying a box is localized and unstructured with some large variations. TRED returns a subsampling only containing points in the localized region with most of the data while maxmin subsampling returns points in the localized region as well as the large variations which incorrectly suggests a larger and more structured time delay embedding. 

\begin{figure}[h]
\includegraphics[width=\columnwidth,keepaspectratio]{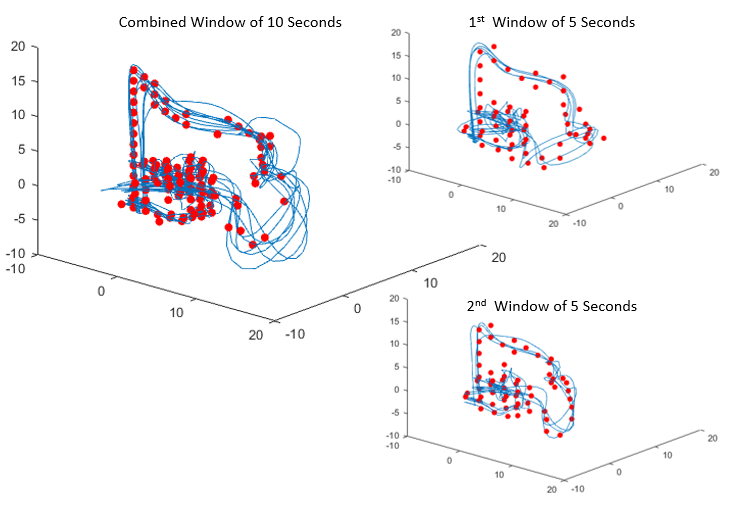}
\caption{The effect of increasing window size on the time delay embedding. The 5 second windows have maximum persistence features of 2.83 and 4.85 (on the right), while the 10 second window has a maximum persistence feature of 1.53 (on the left). This indicates that some of the structure information is lost when considering a longer window size.
}
\label{fig:taueffect}
\end{figure}

The degradation in performance in TRED when increasing the window size is somewhat surprising until inspected closely. It is expected that increasing the window size should increase the number of cycles resulting in a more desirable subsampling. For smaller windows, inconsistent movement can affect the embedding delay $\tau_{d}$, but result in a recognizable structure. For larger windows, the same inconsistent movement has an affect on the $\tau_{d}$ which has a larger effect on the structure by perturbing the pattern which results in less overlap. Figure \ref{fig:taueffect} shows the effect of combining two windows which were classified correctly, but the calculation of $\tau_{d}$ over the combination of the windows differs resulting in a more perturbed embedding.
The same effect on $\tau_{d}$ can be attributed to TRED outperforming maxmin in the 10 sec window as the embedding demonstrates more perturbed, but the outliers are removed in TRED.
By comparing trajectories of the same activity with variable noise, we are able to see a realistic example of TRED effectively capturing the underlying geometric structure of a noisy trajectory as was the focus of Section \ref{sec:synthetic_data}.

We see a similar effect on the performance of maxmin subsampling when increasing the window size, however this effect is quite drastic in the case of maxmin subsampling. The increased window size causes the performance to drop by $ 6 \% $ on average while the performance of TRED only drops $ 3 \%$.
It should be noted that maxmin subsampling is not deterministic and thus running maxmin subsampling again on the same data could produce results which perform either better or worse than the presented data.
The ability to reproduce results is a beneficial characteristic of any algorithm, and as such we see that TRED's deterministic nature is preferred over non-deterministic subsampling.

\subsection{Computational Complexity}

When comparing these two approaches of subsampling for use in activity classification, we find TRED to more reliably extract the underlying geometry and in turn give an effective subsampling. It should be noted that TRED is 4x more efficient than maxmin subsampling. TRED runs at an average rate of 0.0707 seconds per five second window while maxmin subsampling takes an average of 0.2913 seconds. As such TRED's effectiveness in real time subsampling applications is shown even without incorporating the more efficient online insertion and deletion algorithm for streaming applications.
The run time improvement is expected for the type of trajectory data that we are considering as TRED should follow the computational complexity of octrees in this case which is $\mathcal{O}(n)$ considering the depth is a relatively small constant while maxmin subsampling has a computational complexity of $\mathcal{O}(n^2)$.
\section{Street Map Evaluation}
\label{sec:street_map}
We conclude the evaluations by exploring the effectiveness of TRED in extracting the underlying road map of a GPS data set.
We show that TRED is not only effective at filtering out noise (mostly due to the low GPS sampling rate), but also is able to extract a representative map from the raw GPS trajectories. While TRED may not be a domain specific algorithm, we aim to achieve comparable performance with a significant improvement in computational complexity when compared to algorithms written specifically for the purpose of extracting the road network from GPS traces.

\begin{table}[h]
\caption{Overview of Mapping Data set Characteristics 
}
\begin{tabular}{c|c|c|c|c|}
\cline{2-5}
 & Trajectories & \begin{tabular}[c]{@{}c@{}}Trajectory\\ Length\end{tabular} & \begin{tabular}[c]{@{}c@{}}Road\\ Length\end{tabular} & \begin{tabular}[c]{@{}c@{}}Area\\ ($\text{km}^{2}$)\end{tabular} \\ \hline
\multicolumn{1}{|c|}{Athens Large} & 120 & 6,781 km & 2,000 km & 12 x 14 \\ \hline
\multicolumn{1}{|c|}{Athens Small} & 129 & 433 km & 193 km & 2.6 x 6 \\ \hline
\multicolumn{1}{|c|}{Berlin} & 26,831 & 41,116 km & 360 km & 6 x 6 \\ \hline
\multicolumn{1}{|c|}{Chicago} & 889 & 2,869 km & 61 km & 7 x 4.5 \\ \hline
\end{tabular}
\label{table:mapTrackingTable}
\end{table}

\subsection{Streetmap Data Set}
\label{sec:SMDataset}
Ahmed, Karagiorgou, Pfoser, and Wenk \cite{wenkMap} provide a collection of GPS tracking data sets along with state of the art algorithms for extracting a road network from the raw GPS data.
The data sets used in this paper contain both the tracking data set as well as the ground truth with the associated evaluation metrics.
The road networks in these data sets vary in covered area from $15.6 \, km^{2}$ to $168 \, km^{2}$.
We provide an overview of the data collection characteristics for each of the maps and the corresponding characteristics for the ground truth data structures from OpenStreetMap in Table \ref{table:mapTrackingTable}.
We note the methods of data collection vary slightly among the data sets with Athens large and small being collected from the routes of school buses, Berlin collected from a taxi fleet, and Chicago collected from university shuttle buses.
\subsection{Streetmap Reconstruction Evaluation}

We show the results of TRED as evaluated using two metrics presented by Ahmed et. al \cite{Ahmed:2014:LPH:2666310.2666390}, directed Hausdorff distance and path-based distance. 
We now give a short overview of the metrics developed by Ahmed et. al for evaluation.

Let the ground truth graph $G$ have optimal path $G^*$ between two points and the reconstructed graph $H$ have optimal path $H^*$ between two points.
The directed Hausdorff distance can be calculated by taking all points in $G^*$ and finding their closest neighboring point in $H^*$. The maximum distance over a pair of closest point is the Hausdorff distance.
The path-based distance uses the Fr\'echet distance to calculate the metric along $G^*$ and $H^*$ such that the points along the paths are considered in a monotonic manner.
The Fr\'echet distance here is often described as the minimum length leash required for a person to walk along the path $G^*$ while their dog walks along the path $H^*$ from the beginning to end of the path in a monotonic way.

\begin{table*}[]
\caption{Metric Evaluations on TRED Performance as Compared to the Results from \cite{Ahmed:2014:LPH:2666310.2666390}. TRED has Comparable Performance with a Small Fraction of the Computational Time.
}
\label{tab:mapres}
\begin{center}
\begin{tabular}{|c|cccc|cccc|c|}
\hline
\begin{tabular}[c]{@{}c@{}}Generated\\ Map\end{tabular} & \multicolumn{4}{c|}{Path based distance (m)} & \multicolumn{4}{c|}{Directed Hausdorff distance (m)} & Run Time \\ \hline
Berlin                                                  & min      & max       & median      & avg     & min        & max         & median        & avg       & \\ \hline
Ahmed                                                   & 9        & 540       & 66          & 74      & 1          & 219         & 30            & 33        & 30.1 min \\
Ge                                                      & 13       & 808       & 65          & 75      & 4          & 562         & 36            & 37       & -\\
Karagiorgou                                             & 4        & 306       & 28          & 37      & 1          & 232         & 14            & 18        & $>$4 days\\
TRED                                                    & 7        & 626       & 79          & 110     & 1          & 173         & 22            & 24        & 46 sec\\ \hline
Chicago                                                 & min      & max       & median      & avg     & min        & max         & median        & avg       &  \\ \hline
Ahmed                                                   & 7        & 201       & 35          & 42      & 1          & 81          & 14            & 19        & 3.66 min\\
Biagioni                                                & 3        & 71        & 15          & 18      & 2          & 53          & 9             & 11        &-\\
Cao                                                     & 1        & 126       & 24          & 27      & 1          & 78          & 9             & 12        &2.5 days*\\
Davies                                                  & 2        & 92        & 12          & 14      & 2          & 20          & 8             & 7         &14 min*\\
Edelkamp                                                & 1        & 205       & 29          & 37      & 1          & 93          & 8             & 13        &15 min*\\
Ge                                                      & 18       & 346       & 50          & 56      & 7          & 72          & 26            & 28        &-\\
Karagiorgou                                             & 3        & 89        & 15          & 23      & 1          & 48          & 7             & 8         & 15 hr \\
TRED                                                    & 6        & 406       & 62          & 80      & 1          & 173         & 22            & 24        & 4.06 sec \\ \hline
\end{tabular}
\end{center}
\begin{tablenotes}
\item *Runtimes used from \cite{biagioni} due to legacy dependency issues. They are seen to be comparable to runtimes presented by \cite{wenkMap} which are comparable to runtimes of algorithms run on the specified personal computer.
\end{tablenotes}
\end{table*}

In Table \ref{tab:mapres}, we append our results to the results presented by Ahmed et. al \cite{Ahmed:2014:LPH:2666310.2666390} for comparing the success of TRED in extracting the underlying road network from GPS trajectories.
As noted in \cite{Ahmed:2014:LPH:2666310.2666390}, many algorithms failed to produce maps using the GPS traces from both Athens Large and Athens Small.
Similarly, TRED was not evaluated on these datasets due to the small number of trajectories and large portions of the maps with only single trajectories. As a desired property of TRED, those regions with single trajectories are not recovered due to their low trajectoy density. For this reason, the evaluation of TRED on such datasets does not accurately evaluate TRED's ability to extract the underlying geometry of the space which is why those maps were excluded.

\begin{figure}[!b]
\begin{tabular}{cc}
a) Berlin overlay & b) Chicago overlay \\
\includegraphics[width=0.45\columnwidth,keepaspectratio]{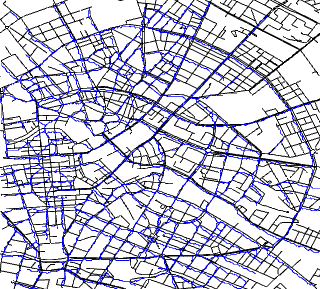} 
&
\includegraphics[width=0.45\columnwidth,keepaspectratio]{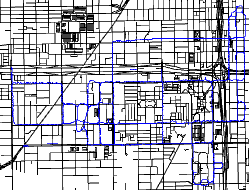}
\end{tabular}
\caption{The TRED reconstruction (blue lines) utilizing the GPS trajectories of vehicles overlaid on the ground truth network (black).}
\label{fig:mapexamples}
\end{figure}

Despite comparing TRED to algorithms which are specifically written to generate road networks from noisy GPS traces, we see TRED is able to perform comparably in both the path based and directed Hausdorff distances.
While performing comparably in the distance metrics for reconstruction accuracy, TRED is able to generate the maps at a vastly reduced computational cost as shown in the last column of Table \ref{tab:mapres}. Figure \ref{fig:mapexamples} shows the output of our method.

We now show how the parameters for TRED align with physical attributes of the underlying maps.
For each of the maps, the parameter \textit{R} which represents the length for the base square that encloses the entire map is defined by the input trajectories.
We empirically determined the maximum depth, \textit{M}, of the quadtree which defined our radius of the finest scale square and the radius offset as $r_1 = R \cdot 2^{-M}$ and $r_2 = (1 - \sqrt{2})\cdot R \cdot 2^{-M}$.
As an analysis on the meaning of our selection of \textit{M}, we compare the radius of finest scale square with the width of a road as our bins should aim to capture a single road.
We show the parameters used to generate the maps in Table \ref{tab:mapinput} as well as the comparison of bin with to the city road width.
It is clear that the selected bin depth produces the bin width closest to the underlying physical city road width. Thus while parameters can be determined empirically, it may be beneficial to utilize known conditions of the space being reconstructed to guide the parameter selection.

\begin{table}[]
\centering
\caption{Output properties of maps with the corresponding parameters used for TRED to produce otimal results. 
}
\label{tab:mapinput}
\begin{tabular}{l|r|r|r|r|r|r|}
\cline{2-7}
                              & Vertices & Edges & \multicolumn{1}{l|}{M} & \multicolumn{1}{l|}{$\tau$} & \multicolumn{1}{l|}{$r_1$ (m)} & \multicolumn{1}{l|}{\begin{tabular}[c]{@{}l@{}}City road\\ width (m)\end{tabular}} \\ \hline
\multicolumn{1}{|r|}{Berlin}  & 6464     & 13530 & 8                      & 35                       & 11                            & 10.7                                                                               \\ \hline
\multicolumn{1}{|r|}{Chicago} & 2794     & 5680  & 8                      & 5                        & 13                            & 11.6                                                                               \\ \hline
\end{tabular}
\end{table}

It should be noted that TRED could be applied to some of the algorithms presented in Table \ref{tab:mapres} or even used as the base structure for an algorithm written specifically for map generation from noisy GPS data. Yet we show the comparison to highlight the effectiveness of our algorithm's success even when compared against tailored and domain specific algorithms. With the effectiveness of a generic TRED algorithm in mind, it would be a valuable candidate for extension into a domain specific algorithm as it currently performs comparably with improved runtime. As such a domain specific extension should either see a superior performance to existing algorithms or a top performance with a superior run time.

We conclude the evaluation of the map construction by noting run time comparisons.
For available source codes, the runtimes on the Berlin and Chicago datasets are appended to Table \ref{tab:mapres} for comparison to TRED. From the runtimes gathered, it is clear that TRED is a far more efficient algorithm while still performing comparably on recovering the underlying road network. 
We see TRED runs in under one minute for each dataset and is well over 30 times faster than even the most efficient algorithm presented. From these performance and runtime comparisons, it is clear that TRED is an extremely efficient algorithm which is able to successfully extract underlying geometry of road networks even without being written as a road extraction algorithm.


\section{Conclusion and Discussion}\label{sec:conclusion}
We propose a method that is able to generate a representative subsampling of points that captures the structure of a trajectory-based density function in space. TRED is robust to sampling frequency both when considering noise due to sampling frequency as well as point density due to sampling frequency. For this reason, TRED is a good candidate for datasets which have noise introduced through sampling techniques as well as inherent inaccuracy of the measurement devices.

We showed TRED is applicable to large data sets which need to be processed in an offline manner as well as streaming data which needs real-time processing.
Our algorithm produces a deterministic datastructure which is desirable in reproducing results and creating guarantees about the performance of the algorithms for some given input data.
The evaluation of TRED on real-time applications, namely activity recognition and street map generation, shows promising results.
Adopting TRED for any spatio-temporal datasets which have an underlying geometry of interest will provide an efficient runtime complexity due to the well understood quadtree data structure while remaining robust to noise.



\section{Acknowledgements}\label{sec:acknowledgements}
This work was supported by the National Science Foundation (NSF) under Award CNS-1552828.

\bibliography{paper_folders/Bibs/Turner_bib}

\begin{thebibliography}{10}

\bibitem{Aanjaneya2011}
Mridul Aanjaneya, Frederic Chazal, Daniel Chen, Marc Glisse, Leonidas~J.
  Guibas, and Dmitriy Morozov.
\newblock Metric graph reconstruction from noisy data.
\newblock In {\em Proceedings of the Twenty-seventh Annual Symposium on
  Computational Geometry}, SoCG '11, pages 37--46, New York, NY, USA, 2011.
  ACM.

\bibitem{agarwal2002}
Pankaj~K. Agarwal, Therese Biedl, Sylvain Lazard, Steve Robbins, Subhash Suri,
  and Sue Whitesides.
\newblock Curvature-constrained shortest paths in a convex polygon.
\newblock {\em SIAM Journal on Computing}, 31(6):1814--1851, 2002.

\bibitem{Ahmed:2014:LPH:2666310.2666390}
Mahmuda Ahmed, Brittany~Terese Fasy, and Carola Wenk.
\newblock Local persistent homology based distance between maps.
\newblock In {\em Proceedings of the 22Nd ACM SIGSPATIAL International
  Conference on Advances in Geographic Information Systems}, SIGSPATIAL '14,
  pages 43--52, New York, NY, USA, 2014. ACM.

\bibitem{wenkMap}
Mahmuda Ahmed, Sophia Karagiorgou, Dieter Pfoser, and Carola Wenk.
\newblock A comparison and evaluation of map construction algorithms.
\newblock {\em CoRR}, abs/1402.5138, 2014.

\bibitem{4059186}
S.~Atev, O.~Masoud, and N.~Papanikolopoulos.
\newblock Learning traffic patterns at intersections by spectral clustering of
  motion trajectories.
\newblock In {\em 2006 IEEE/RSJ International Conference on Intelligent Robots
  and Systems}, pages 4851--4856, Oct 2006.

\bibitem{biagioni}
James Biagioni and Jakob Eriksson.
\newblock Inferring road maps from global positioning system traces: Survey and
  comparative evaluation.
\newblock {\em Transportation Research Record}, 2291(1):61--71, 2012.

\bibitem{CHAKER2017266}
Rima Chaker, Zaher~Al Aghbari, and Imran~N. Junejo.
\newblock Social network model for crowd anomaly detection and localization.
\newblock {\em Pattern Recognition}, 61:266 -- 281, 2017.

\bibitem{chandola2009anomaly}
Varun Chandola, Arindam Banerjee, and Vipin Kumar.
\newblock Anomaly detection: A survey.
\newblock {\em ACM computing surveys (CSUR)}, 41(3):15, 2009.

\bibitem{Chen2010}
Daniel Chen, Leonidas~J. Guibas, John Hershberger, and Jian Sun.
\newblock Road network reconstruction for organizing paths.
\newblock In {\em Proceedings of the Twenty-first Annual ACM-SIAM Symposium on
  Discrete Algorithms}, SODA '10, pages 1309--1320, Philadelphia, PA, USA,
  2010. Society for Industrial and Applied Mathematics.

\bibitem{Chen:2005:RFS:1066157.1066213}
Lei Chen, M.~Tamer \"{O}zsu, and Vincent Oria.
\newblock Robust and fast similarity search for moving object trajectories.
\newblock In {\em Proceedings of the 2005 ACM SIGMOD International Conference
  on Management of Data}, SIGMOD '05, pages 491--502, New York, NY, USA, 2005.
  ACM.

\bibitem{Chentrajectories}
Zaiben Chen, Heng~Tao Shen, Xiaofang Zhou, Yu~Zheng, and Xing Xie.
\newblock Searching trajectories by locations: An efficiency study.
\newblock SIGMOD 2010, June 2010.
\newblock SIGMOD 2010.

\bibitem{de2004topological}
Vin De~Silva and Gunnar~E Carlsson.
\newblock Topological estimation using witness complexes.
\newblock {\em SPBG}, 4:157--166, 2004.

\bibitem{alireza_TDE}
A.~Dirafzoon, N.~Lokare, and E.~Lobaton.
\newblock Action classification from motion capture data using topological data
  analysis.
\newblock In {\em 2016 IEEE Global Conference on Signal and Information
  Processing (GlobalSIP)}, pages 1260--1264, Dec 2016.

\bibitem{DBSCAN}
Martin Ester, Hans-Peter Kriegel, J\"{o}rg Sander, and Xiaowei Xu.
\newblock A density-based algorithm for discovering clusters a density-based
  algorithm for discovering clusters in large spatial databases with noise.
\newblock In {\em Proceedings of the Second International Conference on
  Knowledge Discovery and Data Mining}, KDD'96, pages 226--231. AAAI Press,
  1996.

\bibitem{Finkel}
R.~A. Finkel and J.~L. Bentley.
\newblock Quad trees a data structure for retrieval on composite keys.
\newblock {\em Acta Informatica}, 4(1):1--9, Mar 1974.

\bibitem{Wilfong}
Steven Fortune and Gordon Wilfong.
\newblock Planning constrained motion.
\newblock {\em Annals of Mathematics and Artificial Intelligence}, 3(1):21--82,
  Mar 1991.

\bibitem{Lobaton2016}
Q.~Ge and E.~Lobaton.
\newblock Consensus-based image segmentation via topological persistence.
\newblock In {\em 2016 IEEE Conference on Computer Vision and Pattern
  Recognition Workshops (CVPRW)}, pages 1050--1057, June 2016.

\bibitem{Ge2020}
Qian Ge, Turner Richmond, Boxuan Zhong, Thomas~M. Marchitto, and Edgar~J.
  Lobaton.
\newblock Enhancing the morphological segmentation of microscopic fossils
  through localized topology-aware edge detection.
\newblock {\em Autonomous Robots}, 45(5):709--723, November 2020.

\bibitem{TRACLUS}
Jae-Gil Lee, Jiawei Han, and Kyu-Young Whang.
\newblock Trajectory clustering: A partition-and-group framework.
\newblock In {\em Proceedings of the 2007 ACM SIGMOD International Conference
  on Management of Data}, SIGMOD '07, pages 593--604, New York, NY, USA, 2007.
  ACM.

\bibitem{li2006coarse}
Xi~Li, Weiming Hu, and Wei Hu.
\newblock A coarse-to-fine strategy for vehicle motion trajectory clustering.
\newblock In {\em Pattern Recognition, 2006. ICPR 2006. 18th International
  Conference on}, volume~1, pages 591--594. IEEE, 2006.

\bibitem{li2007traffic}
Xiaolei Li, Jiawei Han, Jae-Gil Lee, and Hector Gonzalez.
\newblock Traffic density-based discovery of hot routes in road networks.
\newblock In {\em International Symposium on Spatial and Temporal Databases},
  pages 441--459. Springer, 2007.

\bibitem{Li2010}
Zhenhui Li, Jae-Gil Lee, Xiaolei Li, and Jiawei Han.
\newblock Incremental clustering for trajectories.
\newblock In {\em Proceedings of the 15th International Conference on Database
  Systems for Advanced Applications - Volume Part II}, DASFAA'10, pages 32--46,
  Berlin, Heidelberg, 2010. Springer-Verlag.

\bibitem{namitaInventions}
Namita Lokare, Boxuan Zhong, and Edgar Lobaton.
\newblock Activity-aware physiological response prediction using wearable
  sensors.
\newblock {\em Inventions}, 2(4), 2017.

\bibitem{mahadevan2010anomaly}
Vijay Mahadevan, Weixin Li, Viral Bhalodia, and Nuno Vasconcelos.
\newblock Anomaly detection in crowded scenes.
\newblock In {\em Computer Vision and Pattern Recognition (CVPR), 2010 IEEE
  Conference on}, pages 1975--1981. IEEE, 2010.

\bibitem{MAKRA20112630}
László Makra, István Matyasovszky, Zoltán Guba, Kostas Karatzas, and Pia
  Anttila.
\newblock Monitoring the long-range transport effects on urban pm10 levels
  using 3d clusters of backward trajectories.
\newblock {\em Atmospheric Environment}, 45(16):2630 -- 2641, 2011.

\bibitem{powell2001accuracy}
Mark~D Powell and Sim~D Aberson.
\newblock Accuracy of united states tropical cyclone landfall forecasts in the
  atlantic basin (1976--2000).
\newblock {\em Bulletin of the American Meteorological Society},
  82(12):2749--2768, 2001.

\bibitem{EUSIPCO}
T.~Richmond, N.~Lokare, and E.~Lobaton.
\newblock Robust trajectory-based density estimation for geometric structure
  recovery.
\newblock In {\em 2017 25th European Signal Processing Conference (EUSIPCO)},
  pages 1210--1204, Aug 2017.

\bibitem{wang2006learning}
Xiaogang Wang, Kinh Tieu, and Eric Grimson.
\newblock Learning semantic scene models by trajectory analysis.
\newblock In {\em European conference on computer vision}, pages 110--123.
  Springer, 2006.

\bibitem{wisdom2004spatial}
Michael~J Wisdom, Norman~J Cimon, Bruce~K Johnson, Edward~O Garton, and
  Jack~Ward Thomas.
\newblock Spatial partitioning by mule deer and elk in relation to traffic.
\newblock In {\em In: Transactions of the 69th North American Wildlife and
  Natural Resources Conference: 509-530}, 2004.

\bibitem{BSNTC}
Y.~Zhang and D.~Pi.
\newblock A trajectory clustering algorithm based on symmetric neighborhood.
\newblock In {\em 2009 WRI World Congress on Computer Science and Information
  Engineering}, volume~3, pages 640--645, March 2009.

\bibitem{Zimek:2013:SEE:2487575.2487676}
Arthur Zimek, Matthew Gaudet, Ricardo~J.G.B. Campello, and J\"{o}rg Sander.
\newblock Subsampling for efficient and effective unsupervised outlier
  detection ensembles.
\newblock In {\em Proceedings of the 19th ACM SIGKDD International Conference
  on Knowledge Discovery and Data Mining}, KDD '13, pages 428--436, New York,
  NY, USA, 2013. ACM.

\end{thebibliography}
\bibliographystyle{plain}
\clearpage
\appendix
\section{Proof of Count Bounds Theorem \ref{thm:boundedCt}}
\label{sec:curv_proof}
\setcounter{equation}{0}
\setcounter{figure}{0}

This proof will built on the theoretical framework developed by Agarwal et al.\cite{agarwal2002} for the analysis of curvature-constrained shortest paths in convex polygons, so we introduce some necessary notation. For a trajectory to be \textit{moderate} it cannot violate its maximum curvature constraint. Let $P$ be a closed convex polygon with boundary $\partial P$. A trajectory whose trace is entirely contained within $P$ is called \textit{free}. If a trajectory is both moderate and free, then we will refer to it as \textit{feasible}. Pockets are constructed from $P$ and the boundary of a ball $\partial B_{r}$ whose radius $r = \frac{1}{\kappa_{max}}$. Let $\partial B_{r}$ intersect $\partial P$ in at least two locations. Consider two consecutive intersection locations on $\partial B_{r}$ where the arc joining the points with length less than $\frac{1}{\kappa_{max}} \cdot \pi$ lies inside $P$. We call these points $A$ and $B$. If the arc joining $A$ and $B$ is clockwise and the turning angle of $\partial P[A,B]$ in the clockwise direction is less than $\pi$, then we define the open region bounded by $\partial B_r[A,B]$ and $\partial P[A,B]$ as a pocket. A pocket is defined in the same manner for a counter clockwise arc $\partial B_{r}[A,B]$ contained within $P$. Let a trajectory \textit{enter} or \textit{escape} a pocket when it crosses the boundary of the pocket defined by $\partial B_{r}[A,B]$. The following results will be essential for our proof:

\begin{lemma} \label{Lem:Pocket}
 If a feasible trajectory enters the interior of a pocket, then it cannot escape the pocket (\cite{agarwal2002}, Lemma 2.7).
\end{lemma}

\begin{figure} 
  \includegraphics[width=\linewidth]{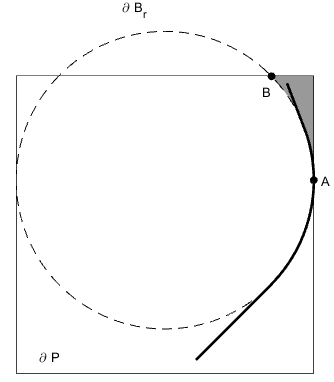}
  \centering
  \caption{Pocket Construction}
  \label{fig:pockets}
\end{figure}

Without loss of generality, let the maximum curvature of $\gamma$ be 1. We drop the variable $x$ from our notation since all sets will be centered around this point. We utilize the following inclusion relationship throughout the proof:
\begin{equation} \label{eqn:SetInclusion}
    \Bone \subset \Sone \subset \Btwo \subset \Stwo.
\end{equation}
We begin by showing the right hand side of Equation \ref{eq:thm1} from the main article holds true. In particular, the steps below show that $C_{\Sone,\Stwo} \leq C_{\Sone,\Btwo} \leq C_{\Btwo}$, where $C_{\Sone,\Btwo}$ is constructed in a similar way to $C_{\Sone,\Stwo}$ except that the identification is done by using the segments within $\Btwo$.

\vspace{0.1in}
\noindent {\bf [Step 1] The bound $C_{\Sone,\Btwo} \leq C_\Btwo$ holds:} Let us define $A_{\Sone,\Btwo} = A_{\Sone}/\sim$ which has as elements the intervals corresponding to segments in $A_{\Sone}$ that are identified using segments corresponding to $A_{\Btwo}$. There is a clear injective map between $A_{\Sone,\Btwo}$ and $A_\Btwo$ (i.e., map the elements in $A_{\Sone,\Btwo}$ to the interval in $A_\Btwo$ that is used on its identification). Hence, we get $C_{\Sone,\Btwo} \leq C_\Btwo$.

\vspace{0.1in}
\noindent {\bf [Step 2] The bound $C_{\Sone,\Stwo} \leq C_{\Sone,\Btwo}$ holds:} We proceed by showing that there is an injective mapping $i: A_{\Sone,\Stwo} \to A_{\Sone,\Btwo}$. Let us take an element $p \in A_{\Sone,\Stwo}$, then this element was formed by taking a subset of intervals $\{[a_j^{(p)},b_j^{(p)}]\}_j\subset\Sone$ and identifying them using a single interval $[c^{(p)},d^{(p)}] \subset \Stwo$. There is also an element $q \in A_{\Sone,\Btwo}$ associated with the interval $[a_1^{(p)},b_1^{(p)}]$ (i.e., the element formed by identifying the set of components that include $[a_1^{(p)},b_1^{(p)}]$). We also have a single interval $[e^{(q)},f^{(q)}] \subset \Btwo$ associated with $q$. We let $i(p)=q$. In order to show that this map is injective, we assume that there exist two elements $p_1$ and $p_2 \in A_{\Sone,\Stwo}$ that map to the same $q$, and show that $p_1 = p_2$. By contradiction, we assume that $p_1$ and $p_2$ are different. In this case, we must have that $[c^{(p_1)},d^{(p_1)}]\neq [c^{(p_2)},d^{(p_2)}]$. If not, then all the segments that each $[c^{(p_k)},d^{(p_k)}]$ identify would be the same, hence making $p_1 = p_2$. Furthermore, remember that the intervals in $A_\Stwo$ are disjoint so $[c^{(p_1)},d^{(p_1)}] \cap [c^{(p_2)},d^{(p_2)}] = \emptyset$. However, since $[e^{(q)},f^{(q)}] \subset [c^{(p_k)},d^{(p_k)}]$ (due to $\Btwo \subset \Stwo$) and $[e^{(q)},f^{(q)}] \neq \emptyset$ (due to the fact that it is associated with $q$) then we have that $[c^{(p_1)},d^{(p_1)}] \cap [c^{(p_2)},d^{(p_2)}] \neq \emptyset$, which is a contradiction. Hence, $p_1 = p_2$ which implies that the mapping is injective and so $C_{\Sone,\Stwo} \leq C_{\Sone,\Btwo}$.

\vspace{0.1in}
By combining steps 1 and 2, we conclude that
\begin{equation}
  C_{\Sone,\Stwo} \leq C_\Btwo.
\end{equation}
Next, we show that $C_\Bone \leq C_{\Sone,\Stwo}$. As done previously, we can show this by constructing an injective mapping $j: A_{\Bone} \to A_{\Sone,\Stwo}$. For any element $p \in A_\Bone$ associated to the interval $[a,b] \subset \Bone$, we can find an interval $[c,d] \subset \Sone$ that contains $[a,b]$ (since $\Bone \subset \Sone$), which in turn is identified to an element $q \in A_{\Sone,\Stwo}$. We want to show that the mapping $j(p) \to q$ is injective. Hence, we want to show that if $j(p_1)= q = j(p_2)$ then $[a_1,b_1] = [a_2,b_2]$. By contradiction, if this was not the case then we would have a segment of the trajectory $\gamma$ that is contained in $\Stwo$ and that escaped and entered $\Bone$ while remaining in $\Stwo$. It needs to escape and enter $\Bone$ in order to form two different components in $\Bone$, and it needs to remain within $\Stwo$ in order to be identified as a single element in $A_{\Sone,\Stwo}$. We will show that this is not possible.

If the trajectory segment in question exists then it either escapes and enters $\Bone$ by remaining in $\Btwo$, or also leaves $\Btwo$ while remaining in $\Stwo$. The later case means that there is a subsegment that escapes and enters $\Btwo$ while entering and escaping a pocket of $\Stwo-\Btwo$ (see Figure \ref{fig:pockets} for an illustration). The following steps show that neither of these cases can happen.


\vspace{0.1in}
\noindent {\bf [Step 3] There cannot exist a trajectory segment of $\gamma$ that escapes and enters $\Bone$ and remains entirely in $\Btwo$:} For a trajectory segment to escape $\Bone$ and reenter while staying in $\Btwo$, it would mean that it would have to curve beyond its curvature limits since $r_2 \leq 1/\kappa_{max}$.

\vspace{0.1in}
\noindent {\bf [Step 4] There cannot exist a trajectory segment of $\gamma$ that leaves and enters $\Btwo$ by entering and escaping a pocket of $\Stwo-\Btwo$:} Such a trajectory segment cannot travel between pockets of $\Stwo-\Btwo$ because the set is four disconnected subsets of $\Stwo$. Since the trajectory must escape $\Btwo$ but remain in $\Stwo$, it must enter one pocket of $\Stwo-\Btwo$ and reenter $\Btwo$ (i.e., escape the pocket). By Lemma 1, a feasible trajectory that enters a pocket from $\Btwo$ cannot reenter through $\Btwo$ (i.e., escape). That is, such a trajectory segment is not possible.

\vspace{0.1in}
Hence, by steps 3 and 4, we must have that $[a_1,b_1] = [a_2,b_2]$, $j$ is injective, and so
\begin{equation}
  C_\Bone \leq C_{\Sone,\Stwo}.
\end{equation}
This completes the proof.

\begin{figure} 
  \includegraphics[width=\linewidth]{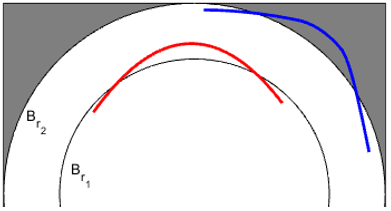}
  \centering
  \caption{$B_{r_{2}}$ is shown as filled in white circle. The four disjoint pockets defined by $S_{r_{2}}-B_{r_{2}}$ are shown as the gray regions. $\Bone$ is depicted with a white circle and black boundary. The red line shows the type of trajectory that we show cannot exist in Step 3. The blue line shows the type of trajectory that we show cannot exist in Step 4.}
  \label{fig:pockets}
\end{figure}

\section{Proof of Stability Theorem \ref{thm:stability}}
\label{sec:stab_proof}

Figure \ref{fig:UnstableSquareCount} from the main article provides a constructive proof showing that $C_{S_r}$ is an unstable count. Furthermore, if $C_{B_r}$ is stable for $r<\frac{1}{\kappa_{max}}$ then by Theorem \ref{thm:boundedCt} from the main article, it is immediate that $C_{\Sone,\Stwo}$ has an upper bound which is controlled by $C_{\Btwo}$. Hence, it follows that $C_{\Sone,\Stwo}$ would be stable for $r_2 < \frac{1}{\kappa_{max}}$. Therefore, the rest of the section focuses on proving that $C_{B_r}$ is stable for $r<\frac{1}{\kappa_{max}}$.

Since we are only considering small $\epsilon$-perturbation, we only need to consider the segments of $\gamma$ that are in $B_{r+\epsilon}$. Anything outside this ball cannot affect the count. Since each one of these segments has a fixed length, and perturbations cannot increase its length indefinitely since they are small (i.e., a perturbed curve cannot wrap around itself indefinitely otherwise it would violate the curvature constraints), then we only need to show that a segment of finite length in $B_{r+\epsilon}$ can only generate a finite count. Alternatively, such segment can only enter and escape the ball $B_r$ a finite number of times.

Without loss of generality, let us assume that the segment starts in $B_r$ and escapes at location $y_1$ and reenters $B_r$ through $y_2$. We show that the distance between $y_1$ and $y_2$ has a lower bounded and hence having an infinite number of entry and exit events is impossible since the segment has finite length.

\begin{figure} 
  \includegraphics[width=\linewidth]{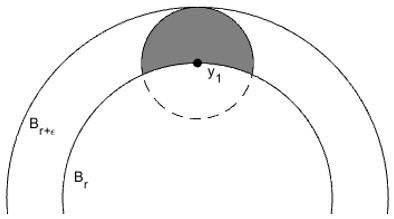}
  \centering
  \caption{Reachable set for a trajectory escaping $B_r$ through $y_1$. Only trajectories within the ball $B_{\epsilon}(y_1)$ are shown.}
  \label{fig:reachable}
\end{figure}

Using the Fortune and Wilfong's algorithm \cite{Wilfong}, we can find the reachable set for the trajectory escaping through $y_1$. Figure \ref{fig:reachable} shows the reachable set restricted by curvature and trajectories entirely within $B_{\epsilon}(y_1)$. Note that none of these trajectory has reentered $B_r$ yet, and they all have length greater or equal to $\epsilon$. Hence, this implies that there can only be a finite number of escape and entry points, and the count is bounded and stable. This concludes the proof.

\end{document}